\documentclass[rmp,aps,twocolumn]{revtex4}
\usepackage{graphicx} % Style for including EPS files
\usepackage{amsmath}  % AMS style for formulas etc. 
\usepackage[]{caption} % Aligned captions instead of centerede

\DeclareMathOperator{\Erfc}{Erfc}
\DeclareMathOperator{\Erf}{Erf}
\bibpunct{(}{)}{;}{a}{,}{,}

\begin{document}
\title{The binding dynamics of tropomyosin on actin} 
\author{Andrej Vilfan}
\affiliation{Cavendish Laboratory, Madingley Road, Cambridge CB3 0HE, UK}
\email{av242@phy.cam.ac.uk}

\date{30/10/2001}

\begin{abstract}We discuss a theoretical model for the cooperative binding
  dynamics of tropomyosin  to actin filaments.  Tropomyosin binds
  to actin by occupying seven consecutive monomers.  The model includes
  a strong attraction between attached tropomyosin molecules.  We
  start with an empty lattice and show that the binding goes through
  several stages. The first stage represents fast initial binding and
  leaves many small vacancies between blocks of bound molecules.  In
  the second stage the vacancies annihilate slowly as
  tropomyosin molecules detach and re-attach.  Finally the system
  approaches equilibrium. Using a grain-growth model and a
  diffusion-coagulation model we give analytical approximations for
  the vacancy density in all regimes.
\end{abstract}

\pacs{}

\keywords{Nucleation; Annealing; Reaction-diffusion models; Random
sequential adsorption (RSA)} 

\maketitle

\section*{INTRODUCTION}

\begin{figure}
\begin{center}
\includegraphics{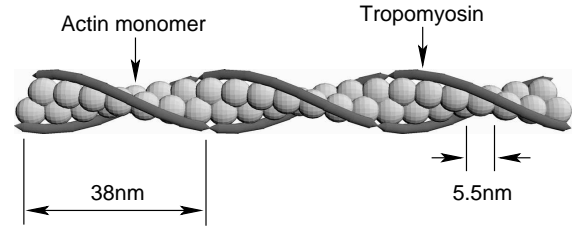}
\end{center}
\caption{Tropomyosin (TM) binds on actin polymers on each
side of the double helix. One TM molecule thereby occupies 7
actin monomers.}
\label{fig_3d}
\end{figure}

Tropomyosin (TM) is the protein that plays the key role in the
regulation of muscle contraction \citep{lehrer98}.  It comes in a
rod-like form and binds to the groove each side of double-helical actin
filaments (Fig.~\ref{fig_3d}).  Actin filaments give the cytoskeleton
its mechanical stability and serve as tracks for motor proteins from
the myosin family, especially in muscle cells.  In muscle each
tropomyosin molecule covers 7 actin monomers
\citep{mclachlan76,hitchcock-degregori90}. Other tropomyosins covering
6 and 5 monomers exist as well (reviewed by
\citet{hitchcock-degregori94}), but for specificity we will focus on
TM covering 7 actin subunits.  Together with the associated protein
troponin, TM on actin can switch between two laterally shifted
conformations \citep{vibert97,lehman2000}.  The equilibrium between
these two conformations is strongly influenced by the concentration of
$\rm Ca^{2+}$ ions. As one conformation obstructs the strong binding of
myosin to actin \citep{geeves94,lehrer98}, this provides the mechanism
of calcium-mediated activation of myosin in skeletal muscle cells.
The regulation of myosin activity has been widely studied using
Actin-tropomyosin filaments assembled in vitro
\citep{fraser95,gordon97}.  Other functions of TM might include the
stabilization of actin against fragmentation and actin assembly
\citep{lazarides76,weigt91}.

The binding of tropomyosin to actin has been studied in equilibrium
\citep{wegner79,hill92} as well as dynamically
\citep{wegner88,weigt91} and has been found to be highly cooperative
\citep{wegner79,hill92}.  But the way that tropomyosin binds to actin
is not simple.  As each TM molecule occupies 7 actin monomers, an
obvious problem is that sometimes a gap of 1-6 monomers, too short for
another TM molecule to be inserted, can remain on actin.  It has been
suggested \citep{weigt91} that these gaps could play an important role
by regulating the binding of $\alpha$-actinin and the fragmentation
kinetics of actin.

\begin{figure}
\begin{center}
\includegraphics{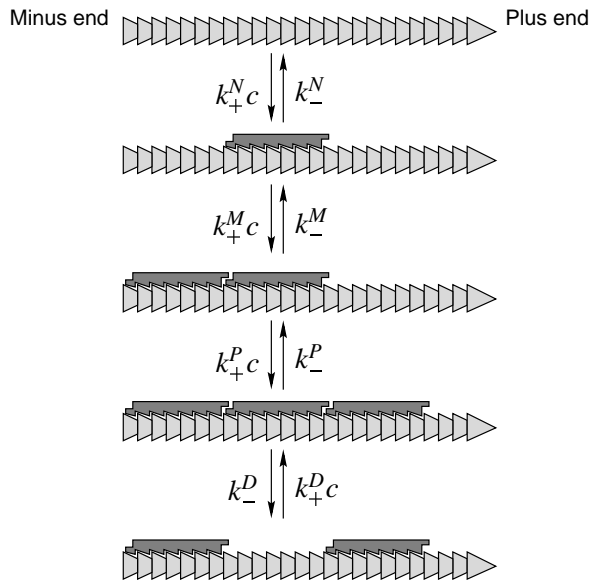}
\end{center}
\caption{Reaction scheme for all binding and
unbinding processes of TM on actin.  The binding (unbinding) rate for
an isolated site is $k_+^Nc$ ($k_-^N$), at the ``$-$'' side of another bound
TM molecule $k_+^Mc$ ($k_-^M$), at the ``$+$'' side $k_+^Pc$ ($k_-^P$), and
between two bound TM molecules $k_+^Dc$ ($k_-^D$). $c$ denotes the
concentration of TM in solution.}
\label{fig_scheme}
\end{figure}

In this Article, we study a theoretical model for the dynamics of
binding of TM to actin based on the reaction kinetics shown in
Fig.~\ref{fig_scheme}, first proposed by Wegner and coworkers
\citep{wegner79,weigt91}.  The equilibrium distribution of bound
molecules and vacancies has been known exactly for a long time
\citep{wegner79} and represents a special case of a more general model
solved by \citet{mcghee74}.  However, as we have previously shown in
the context of kinesin on microtubules \citep{vilfan2001a,vilfan2001b},
the time necessary to reach the equilibrium state can be very long.
There are many similarities between the binding of dimeric kinesin on
microtubules and TM on actin.  In both situations there is an initial
period in which all gaps that are wide enough to accept a new molecule
get filled.  The dynamics within this period has been directly
experimentally observed using fluorescent techniques
\citep{weigt91}. After the initial period, only vacancies stretching
over 1 to 6 actin monomers, too small for another TM molecule to fit
in, remain.  They get healed out in a much slower annealing process in
which bound molecules detach and others re-attach at other places.
This process takes place on a much slower time scale.  The annealing
process is finished when the system reaches the equilibrium state.
Despite the similarities with the binding of kinesin onto
microtubules, there are also important differences:
\begin{enumerate}
\item The TM molecules occupy 7 instead of two binding sites.
\item Kinesin dimers can bind with one head as well as with two, while
TM must either bind all sites or be fully detached.
\item The interaction between bound tropomyosin molecules is much
stronger.
\end{enumerate}
In our theoretical analysis we make reasonable simplifications
justified by experimental observations and then use results from
diffusion-annihilation models to find an analytical solution for the
dynamics in our model.  The central result is the prediction of the
number of defects (gaps between contiguous blocks of tropomyosin) as
a function of time for a given solution concentration.

The fact that a tropomyosin molecule occupies 7 binding sites makes it
especially interesting from the theoretical point of view. The size 7 lies
between the simple monomeric or dimeric and continuous models.  We have shown
previously that the relaxational behavior of the dimer adsorption-desorption
model can be explained by a mapping onto the $A+A\to 0$ diffusion-annihilation
model \citep{vilfan2001a}.  The gap density decays with a power-law
$\propto t^{-1/2}$ before it approaches the final equilibrium value.  This
behavior remains qualitatively valid even if there is an attractive
interaction between adsorbed dimers.  If one introduces a strong diffusion of
adsorbed dimers the power-law decay is preceded by a mean-field regime in
which $n_G\propto t^{-1}$ \citep{privman92}. On the other hand,
non-interacting $k$-mers ($k>3$) are always well described with the mean-field
kinetics of the model $kA\to 0$ which predicts a gap density decay $\propto
t^{-1/(k-1)}$ \citep{nielaba92}.  In the continuous limit, also called
continuous sequential adsorption (CSA) or ``car parking problem'' the length
of unoccupied space decays as $\propto 1 / \log t$ in the mean-field
regime \citep{krapivsky94,jin94}.  We will show that our model ($k=7$ with
strong attractive interaction) differs from all models mentioned above and
that it maps onto a $A+A\to \frac56 A$ model, which includes particle
annihilation as well as coagulation.

\section*{DEFINITION OF THE MODEL}

Our model is based on the reaction scheme shown in
Fig.~\ref{fig_scheme} and is defined as follows.  We assume that
there is no interaction between bound TM molecules on the opposite
sides of the actin polymer and therefore describe actin as two
independent one-dimensional binding lattices.  We further assume that
TM molecules can bind to actin only with their whole length, thereby
occupying 7 binding actin monomers.  
The rate at which a TM molecule binds to a group of seven free binding
sites on actin is $k_+^N c$ where $c$ denotes the TM concentration.
The rate for the reverse reaction, i.e., the detachment of a TM
molecule without neighbors, is $k_-^N$.  The rate at which new
TM molecules bind beside already bound ones (we call such sites
single-contiguous binding sites) will be denoted as $k_+^P c$ (the
rate at which the new TM binds on the plus side of a previously
occupied block) and $k_+^M c$ (on the minus side).  The reverse
reaction rates are $k_-^P$ and $k_-^M$.  It is reasonable to assume
that the binding constants are only influenced by nearest neighbors.
Finally the attachment rate in a gap of exactly 7 lattice sites will
be $k_+^D c$ and the detachment rate in the middle of a contiguous
block $k_-^D$.  The principle of detailed balance states that the
ratio between the binding and unbinding rate depends only on the free
energy difference and therefore has the same value regardless whether
a molecule binds on the plus or minus side of an occupied block
\begin{equation}
\frac{k_+^M}{k_-^M}=\frac{k_+^P}{k_-^P}=K= \frac{k_+^N}{k_-^N} \exp\frac
J {k_B T}=  K^N \exp \frac J {k_B T}
= K^N \gamma\;.
\end{equation}
Here $J$ denotes the coupling energy between two TM molecules on
actin and $\gamma=\exp(J/k_BT)$ the cooperativity coefficient
\citep{hill85}.  The coupling energy from both ends is additive,
therefore
\begin{equation}
\frac{k_+^D}{k_-^D}=K^D=\frac{k_+^N}{k_-^N} \exp \frac {2J}
{k_B T}=K^N \gamma^2 \;.
\end{equation}

In the following we will assume a strong interaction between bound
molecules, $\gamma \gg 1$.  The assumption is well justified since the
cooperativity coefficient has been determined to be between $\gamma=600$ and
$\gamma=1000$ by \citet{wegner79} and around $\gamma=100$ by
\citet{hill92}.  The binding affinity of an isolated TM molecule has
been found to be very low \citep{weigt91} and we may assume
\begin{equation}
K^Nc \ll 1\;. \label{eq_singleunstable}
\end{equation}
When discussing the dynamics we will neglect all processes where a TM
molecule detaches in the middle of a contiguous block, setting $k_-^D
\approx 0$.

\section*{RESULTS}

The equilibrium solution of the problem on an infinite lattice has
been known exactly for a long time \citep{mcghee74,wegner79}.  The
number of attached TM molecules per lattice site $n_{\rm TM}$ (which
can assume values between $0$ and $1/7$) is given by the following
implicit equation
\begin{multline}
\label{eq_mcghee}
K^N c = \frac{n_{\rm TM}}{1-7 n_{\rm TM}} \left( \frac{2 (\gamma-1)(1-7 n_{\rm TM})}{(2 \gamma-1) (1-7 n_{\rm TM})+n_{\rm TM}-R}\right) ^6 \\
\times \left(\frac{2(1-7 n_{\rm TM})}{1-8n_{\rm TM}+R}\right)^2
\end{multline}
with
\begin{equation*}
R=\sqrt{(1-8n_{\rm TM})^2+4 \gamma n_{\rm TM} (1-7n_{\rm TM})}\;.
\end{equation*}
The average gap size was determined as
\begin{equation}
\bar g=\frac{2 (\gamma-1)(1-7 n_{\rm TM})}{-(1-6 n_{\rm TM}) +R}\;.
\end{equation}
The fraction of unoccupied lattice sites is given as $n_0^{\rm Eq}=1-7
n_{\rm TM}$ and the gap density as $n_G^{\rm Eq}=n_0^{\rm Eq}/\bar
g$. In the limit of very high TM concentrations and nearly full
coverage, $n_{\rm TM} \to 1/7$, the average gap size becomes $\bar g \approx
1$.

\begin{figure*}
\begin{center}
\includegraphics{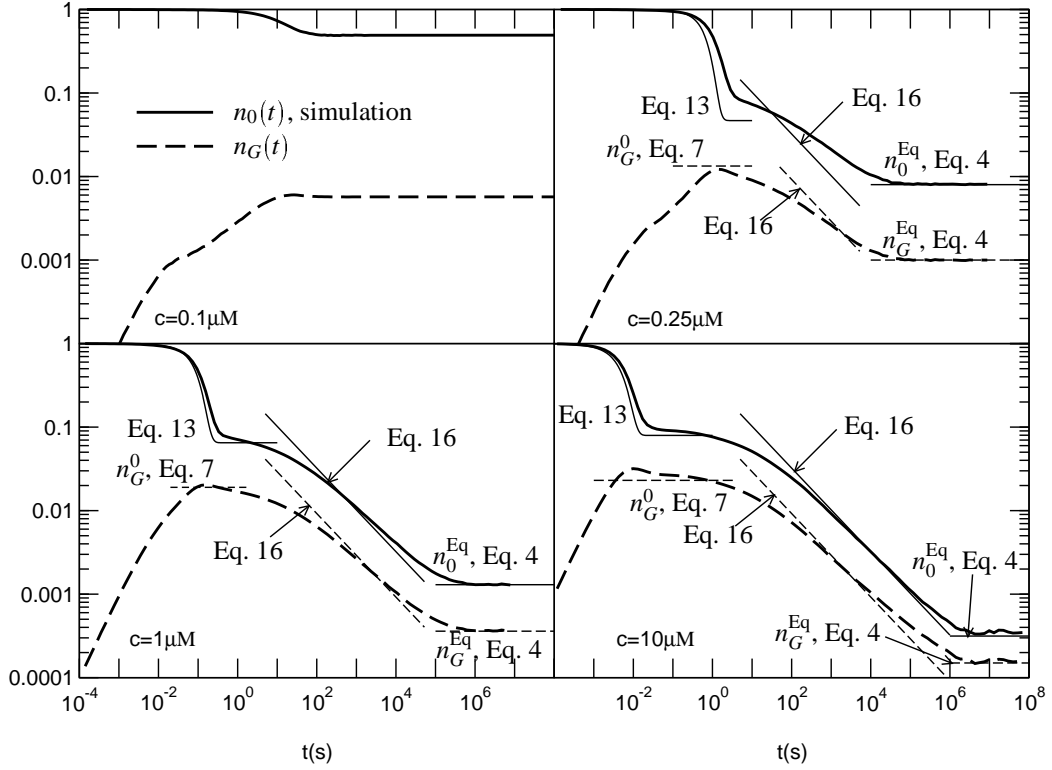}
\end{center}
\caption{Gap density ($n_G(t)$, dashed line) and
fraction of unoccupied binding sites ($n_0(t)\equiv 1-7 n_{\rm
TM}(t)\equiv n_G(t)\bar g(t)$, solid line) as a function of time,
obtained from a stochastic simulation. Parameters: $k_+^N=1.0\,\mu{\rm
M}^{-1}s^{-1}$, $k_+^P=50\,\mu{\rm M}^{-1}s^{-1}$, $k_+^M=3.0\,\mu{\rm
M}^{-1}s^{-1}$, $k_-^N=100\,s^{-1}$ and $\gamma=1000$. The four graphs
show data for different TM concentrations: 0.1, 0.25, 1 and
$10\,\mu{\rm M}$. With these values the effective hopping rate is
$r_{\rm hop}=0.28\,s^{-1}$.  The thin lines show the predictions of
Eq.~\ref{eq_nvt} for the initial phase, Eq.~\ref{eq_ng_plateau}
for the plateau, Eq.~\ref{eq_ngt} for the algebraic regime and
eventually the exact result of Eq.~\ref{eq_mcghee} for the
equilibrium. The agreement between the theoretical calculations and
the simulation becomes very good for concentrations $c \ge 1\,\mu{\rm
M}$.  Simulation data were obtained on a lattice large enough that the 
results correspond to an infinite system ($2^{17}$ sites and periodic 
boundary conditions).}
\label{fig_sim}
\end{figure*}

However, the exact equilibrium configuration is of little value
without the knowledge of the time the system needs to equilibrate
(relaxation time).  Figure \ref{fig_sim} shows the time dependent
filament coverage and gap concentration as obtained from a dynamic
Monte-Carlo simulation for different TM concentrations with realistic
values for the kinetic constants. As the simulation results show, the
dynamics shows an interesting two-stage behavior at high TM
concentrations.  In the first stage, TM molecules bind to actin
wherever there is enough unoccupied space.  The first stage is
followed by a plateau and the second stage in which reordering of
bound TM molecules takes place.  In this second stage gaps that
inevitably remain after the first stage are healed in an annealing
process.  The second stage is finally followed by an exponential
relaxation into the equilibrium state given by Eq.~\ref{eq_mcghee}. The
second stage is much slower than the first one.  We therefore conclude
that in many experimental situations the relevant properties of the
system are actually determined by the dynamic behavior of the model and
not its equilibrium state.

\subsection*{Initial binding}

In this Section we describe the initial dynamics of TM starting with
an empty actin filament.  Since a single TM molecule only weakly binds
to actin (Eq.~\ref{eq_singleunstable}) nucleation is needed to initiate
binding.  For its description we use the approach first developed by
\citet{kolmogorov37} in his `grain growth' model (reviewed by
\citet{evans93}; see also \citep{privman97}).  Kolmogorov's model
assumes that nucleation is sufficiently slower than grain growth,
meaning that grains can grow to large sizes before the growth process
is stopped by another grain.  It therefore neglects fluctuations in
the growth process and assumes a constant growth velocity.  The
nucleation, on the other hand, is described with its full
stochasticity.

\begin{figure}
\begin{center}
\includegraphics{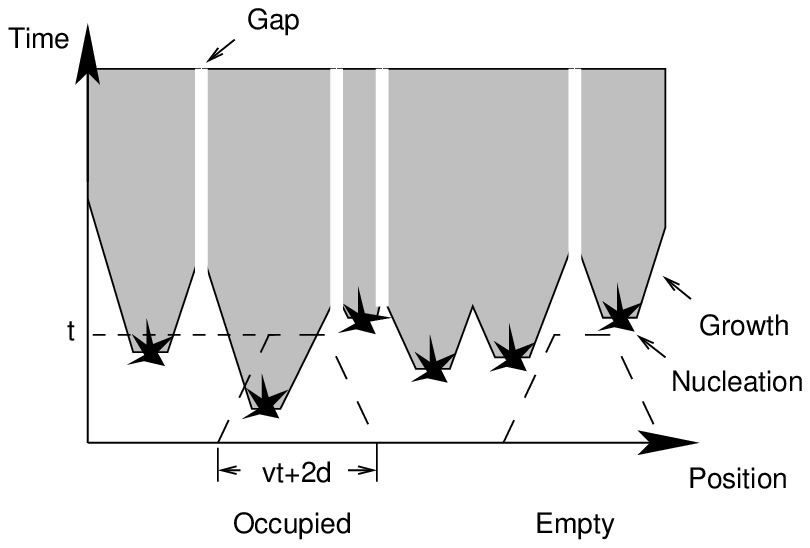}
\end{center}
\caption{Nucleation and growth of tropomyosin (TM)
blocks on actin.  Each time the blocks from two different nuclei meet,
a gap remains in 6 out of 7 cases.  A lattice site at time $t$ can
be considered as empty if nucleation has not taken place in a trapezium
of height $t$, base $vt+2d_0$ and top $2d_0$ where $v$ is the sum of
growth velocities at both ends and $d_0$ the size of a nucleus.}
\label{fig_newnucleus}
\end{figure}

Where the blocks growing from two neighboring nuclei meet, there is a
probability of $\frac 1 7$ that they will match exactly without
leaving a gap (Fig.~\ref{fig_newnucleus}).  In all other cases,
occurring with a total probability of $\frac 6 7$ a gap of 1-6 sites
will remain.  The number of such gaps should therefore equal $\frac 6 7$ of
the total number of nuclei.  All gap sizes between 1 and 6 sites
occur with equal probabilities and the average gap width is 3.5
sites.

We calculate the number of gaps after the initial binding process in
the following way.  For simplicity we assume that the actin
concentration is low enough that the TM concentration in the solution
does not change significantly in time (a generalization for a
time-dependent solution concentration is discussed later).  Then the
nucleation rate $r_n$ and the growth velocity $v$ are constant in
time. The probability that a nucleus will be formed on site $i$ at
time $t$ equals the nucleation rate at that time multiplied by the
probability that the site has not been occupied previously. The latter
equals to the probability that there is no nucleus in a triangle of
height $t$ and base $vt$ in the position-time diagram (akin to the
light cone frequently used to illustrate independent events in the
special theory of relativity) (Fig.~\ref{fig_newnucleus}).  To improve
the accuracy at relatively high nucleation rates, we can take into
account that the nucleus already has a width of $d_0=14$ sites and the
triangle then becomes a trapezium with side lengths $2d_0+v t$ and
$2d_0$ and height $t$.

\begin{figure}
\begin{center}
\includegraphics{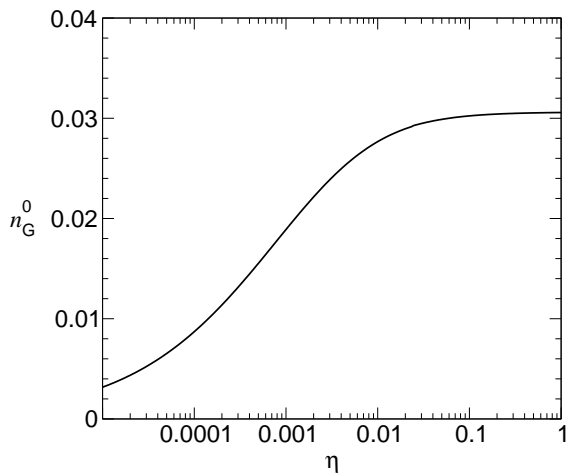}
\end{center}
\caption{The gap density (number of gaps per lattice site) after the
initial binding phase, $n_G^0$ as a function of the parameter
$\eta=r_n/v$, as given by Eq.~(\protect\ref{eq_ng_plateau}).}
\label{fig_eta}
\end{figure}

The probability that no nucleation event takes place within the trapezium is given by the zeroth term of a Poisson distribution with the expectation value 
$\bar N=r_n (vt/2+2d_0)t$, namely  $P_0=\exp(-\bar N)$.  Then the
nucleation rate at time $t$ reads
\begin{equation}
\label{eq_p0}
{\bar r}_n(t)=r_n P_0(t)=r_n e^{-r_n \left( \frac{vt}{2}+2d_0 \right)t}\;.
\end{equation}
The density of gaps per lattice site in the state after initial
attachment is then given as
\begin{align}
\label{eq_ng_plateau}
n_G^0&=\frac 6 7 \int_0^\infty dt\, {\bar r}_n(t)\nonumber \\&=\frac 6 7 \sqrt{\frac{\pi}{2} \eta}\, e^{2d_0^2 \eta} \Erfc\left(d_0
\sqrt{2 \eta} \right)\quad \text{with}\;\eta=\frac{r_n}{v} 
\end{align}
and is plotted in Fig.~\ref{fig_eta}.  Its asymptotic limits are
\begin{align}
n_G^0(\eta)&=\frac 6 {14 d_0}&& \text{for}& \eta &\to \infty \nonumber \\
n_G^0(\eta)&=\frac 6 7 \sqrt{\frac \pi 2 \eta}&& \text{for} & \eta &\to 0\;.
\end{align}

Once we have expressed the gap concentration with the nucleation and
growth rate, we need to express these with the original model
parameters.  The total growth velocity $v$ (sum of growth velocities
at the $+$ and the $-$ end, expressed in lattice sites per time unit)
reads
\begin{equation}
\label{eq_v}
v=7 \left[(k_+^P+k_+^M)c-(k_-^P+k_-^M)\right] \approx 7 (k_+^P+k_+^M)c
\;.
\end{equation}
In this equation we neglected the effect that uncontiguously bound
molecules (which detach after a short time) would have on the growth
by blocking the binding sites.  This approximation is justified as
long as $K^Nc \ll 1$.

If $Kc\gg 1$, two molecules bound beside each other already form a
stable nucleus, while a single one does not
(Eq.~\ref{eq_singleunstable}). The nucleation rate $r_n$ can be obtained as the
attachment rate of the first molecule multiplied by the probability
that a second one binds to its side before the first one detaches and reads
\begin{equation}
\label{eq_rn}
r_n=k_+^N\frac{(k_+^P+k_+^M)c^2}{(k_+^P+k_+^M)c+k_-^N}\;.
\end{equation}
Together with Eq.~\ref{eq_v} we obtain
\begin{equation}
\label{eq_eta}
\eta=\frac{k_+^N c}{7 ((k_+^P+k_+^M)c+k_-^N) }\;.
\end{equation}
In the limit of fast equilibration of isolated TM molecules (i.e., if
the detachment of an isolated molecule is much faster than the
attachment of another one to its side) it simplifies to
\begin{align}
\label{eq_rn2}
\eta&=K^N c/7 &&\mbox{for}&
k_-^P+k_-^M\ll(k_+^P+k_+^M)c&\ll k_-^N\;.
\end{align}
The values of the plateau gap concentration obtained from
Eq.~\ref{eq_ng_plateau} are shown in Fig.~\ref{fig_sim} (short
dashed line).  They show good agreement with the simulation value as
long as the assumption $Kc\gg 1$ is fulfilled.  We can go even further
in our analysis and give an approximation for the time dependent
vacancy concentration.  We can estimate the fraction of empty binding
sites at time $t$ as the probability that no nucleus has reached that
point plus the number of empty sites that remain in gaps between
nuclei (on average 3 per nucleus).  Therefore, we obtain
\begin{align}
\label{eq_nvt}
n_0(t)&\equiv 1 - 7 n_{\rm TM}(t)=\nonumber \\
&= e^{-r_n \left( \frac{vt}{2}+d_0 \right)t} +
3 \int_0^t dt' r_n e^{-r_n \left( \frac{vt'}{2}+2 d_0 \right)t'}=\nonumber \\
&=e^{-r_n \left( \frac{vt}{2}+d_0 \right)t} +3 \sqrt{\frac{\pi r_n}{2 v}} e^{\frac {2 d_0^2 r_n} {v}}
\nonumber\\
& \left[ \Erf \left( \sqrt{\frac{2 r_n}{v}}\left( d_0 +\frac{vt}2 \right) \right) - \Erf \left( \sqrt{\frac{2 r_n}{v}} d_0 \right) \right] \;.
\end{align}
The comparison between the prediction of Eq.~\ref{eq_nvt} and the
simulation result can be seen in Fig.~\ref{fig_sim}.  The
agreement becomes very good at concentrations $c\ge 1\,\mu{\rm M}$.

\subsection*{Annealing of gaps}

\begin{figure}
\begin{center}
\includegraphics{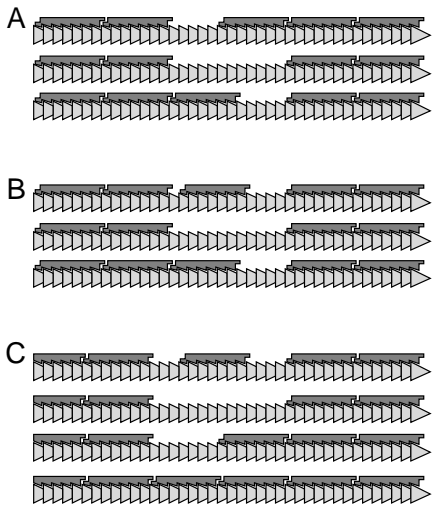}
\end{center}
\caption{Examples of effective reactions that
occur after the detachment and attachment of a tropomyosin molecule:
Diffusive step (A); Pair coagulation $A+A\to A$ (B); Pair annihilation
$A+A\to 0$ (C).}
\label{fig_reactions} 
\end{figure}

By now we have calculated the gap concentration after the initial
binding phase.  What follows is a process on a much longer time-scale
in which molecules at edges of the gaps can detach and re-attach.  If they
re-attach on the other side of a gap, the gap makes a move of 7 sites
in one direction (Fig.~\ref{fig_reactions}a).  The rate of such diffusive steps to each direction
is given as the detachment rate of a TM molecule on e.g.\ the minus
side of a block multiplied by the probability that a molecule
re-attaches on the plus side of the other block before another one
re-attaches to the position where the first one has detached:
\begin{equation}
\label{eq_hop}
r_{\rm hop}=k_-^M \frac {k_+^P}{k_+^P+k_+^M}=\frac 1 K \frac{k_+^P
k_+^M}{k_+^P+k_+^M}= \left( \frac 1 {k_-^P}+\frac 1 {k_-^M}
\right)^{-1} \;.
\end{equation}

If two gaps, each containing 1-6 sites with equal probabilities,
come together, they can either join to a single gap
(Fig.~\ref{fig_reactions}b) or, if their total size exactly fits one
TM molecule, annihilate (Fig.~\ref{fig_reactions}c).  If the original
gap sizes are $g_1$ and $g_2$, then the joined gap has the size
$(g_1+g_2) \mod 7$.  All possible combinations of $g_1$ and $g_2$ are
shown in Table \ref{table1}.
\begin{table}
\begin{center}
\begin{tabular}[c]{cc|cccccc}
\multicolumn{2}{c}{}
&\multicolumn{6}{c}{$g_2$}\\
&& 1 & 2 & 3 & 4 & 5 &6\\
\cline{2-8}
        &1 & 2 & 3 & 4 & 5 & 6 & -\\
        &2 & 3 & 4 & 5 & 6 & - & 1\\
$g_1$   &3 & 4 & 5 & 6 & - & 1 & 2\\
        &4 & 5 & 6 & - & 1 & 2 & 3\\
        &5 & 6 & - & 1 & 2 & 3 & 4\\
        &6 & - & 1 & 2 & 3 & 4 & 5\\
\end{tabular}
\end{center}
\caption{Size of the joint gap after coagulation of gaps with sizes
$g_1$ and $g_2$.  If the initial gap sizes $g_1$ and $g_2$ are randomly
distributed with values between 1 and 6, the probability for
annihilation is $1/6$.  The probability that the joint gap will have a
certain size between 1 and 6 is always $5/36$.}
\label{table1}
\end{table}
This table shows that if the probabilities of both gap sizes $g_1$ and
$g_2$ are equally distributed between $1$ and $6$, the same will hold
for the size of the resulting gap $g_j$. In $1$ out of $6$ cases the
two gaps will annihilate, while they will form a new gap with the same
size distribution in $5$ out of $6$ cases.  Thus we can represent gaps
as particles $A$, hopping randomly along the lattice and joining or
annihilating when two of them meet.  We can therefore map our model to
a diffusion-annihilation model consisting of following reactions
\begin{align*}
A+A&\to A& \text{probability}\; 5/6\\
A+A&\to 0& \text{probability}\; 1/6
\end{align*}

According to \citet{lee94} all diffusion-annihilation models of the
type $2A\to l A$ (or generally $k A\to l A$) belong to the same
universality class. In the asymptotic limit, the ``particle''
concentration can be related to that of the exactly solvable $A+A\to
0$ model \citep{lushnikov87,torney83} and reads
\begin{equation}
n(t)=\frac 2 {2-l} \frac 1 {\sqrt{8 \pi \bar{r}_{\text{hop}} t}}\;.
\end{equation}
In our case we have to set $l=5/6$ and $\bar{r}_{\text{hop}}=7^2
r_{\text{hop}}$.  The second relation results from the fact that in
each diffusive step a gap jums over seven sites.  We finally obtain
\begin{equation}
\label{eq_ngt}
n_G(t)=\frac 6 {49 \sqrt{2 \pi r_{\text{hop}} t}}\quad \text{and}\quad n_0(t)=3.5\, n_G(t)\;.
\end{equation}
Interestingly, the asymptotic particle concentration is independent of
its initial value.  The gap concentration at long times therefore does
not depend on the intermediate gap concentration $n_G^0$. As our
equation reveals, it is even independent of the solution concentration
$c$.

Of course, the mapping to the $A+A\to l A$ model is only valid as long
as the concentration of particles $A$ is well above its equilibrium
value.  When these come closer, events of pair creation like
$A\to A+A$ become relevant.  One can therefore estimate the
equilibration time as the time when the vacancy concentration
determined by Eq.~\ref{eq_ngt} becomes equal to the equilibrium
concentration from Eq.~\ref{eq_mcghee}.  This consideration is
only valid if the filaments are long enough that even in equilibrium a
considerable number of gaps remains.  Otherwise the equilibration time
can be estimated as the time in which the gap concentration drops
below one per filament.

Figure \ref{fig_sim} also shows the comparison between the result of
Eq.~\ref{eq_ngt} and the simulation result.  The results agree well
for high concentrations where the intermediate plateau and the final
equilibrium are far apart.  At very high concentrations (not shown, as
they would be beyond experimental relevance), there would be an
additional power-law regime between the $t^{-1/2}$ law shown here and
the final equilibrium.  In that regime processes of the type $A\to
A+A$ become relevant, but not yet processes of the type $0\to A+A$.
The model then becomes equivalent to the non-interacting 7-mer
disposition model in which the vacancy density decays according to the
mean-field law $\propto t^{-1/6}$ \citep{nielaba92}.  But with
experimentally relevant parameters one can see only a slight remnant
of this regime.

\begin{table*}
\begin{equation*}
\begin{array}[c]{lr@{\qquad}l@{\qquad}ll@{\qquad}ll}
\hline
&&\text{Fig.~\ref{fig_sim}}&\multicolumn{2}{c}{\text{\citep{weigt91}\footnotemark[1]}}&\multicolumn{2}{c}{\text{\citep{hill92}\footnotemark[2]}}\\
&&&\multicolumn{2}{c}{100\,{\rm mM\; KCl}}&60\,{\rm mM\; KCl}&300\,{\rm mM\; KCl}\\
\hline
&K^N (\mu{\rm M}^{-1})&{\bf  0.01 }& 0.005 & 0.026& 0.017&0.00058\\ 
&k_+^P+k_+^M (\mu{\rm M}^{-1}s^{-1}) &{\bf 53}& 87 & 38 &&\\
&k_-^N (s^{-1}) &{\bf 100 }& 100 & 100 &&\\
&\gamma&{\bf 1000}&1000& 1000&90&138\\
&k_+^M(\mu{\rm M}^{-1})&{\bf 3.0}& 3.0 & 3.0 &&\\
\hline
&r_{\rm hop}\,({\rm s}^{-1})&{\bf  0.28 }&0.58&0.11 &&\\
&t_{0.001}\,({\rm s})&{\bf 8400}&4100&22000&&\\
&t_{0.01}\,({\rm s})&{\bf 84}&41&220&&\\
\hline
c=0.1\mu {\rm M}& 
n_G^{Eq}&{\bf }&  & 0.00095&&\\
&{\bar g}&{\bf 86}& 1000 &  7.8&510&17000\\
&\eta&{\bf }&&0.00036&&\\
&n_G^0&{\bf }&& 0.014&&\\
&t^{\rm Eq}\,({\rm s})&{\bf }&&24000&&\\
\hline
c=1\mu {\rm M}& 
n_G^{Eq}&{\bf 0.00036}& 0.00055 & 0.00024&&\\
&{\bar g}&{\bf 3.6}& 4.9& 2.7& 13&1600\\
&\eta&{\bf 0.00093}&0.00038&0.0027&&\\
&n_G^0&{\bf 0.019}&0.014&0.023&&\\
&t^{\rm Eq}\,({\rm s})&{\bf 63000}&14000& 390000&&\\
\hline
c=10\mu {\rm M}& 
n_G^{Eq}&{\bf 0.00015}&0.00019 & 0.00012 &0.0032&\\
&{\bar g}&{\bf 2.1 }& 2.3 & 1.8 & 3.1 & 58\\
&\eta&{\bf 0.0023}&0.00074&0.0077&&\\
&n_G^0&{\bf 0.023}&0.017& 0.027&&\\
&t^{\rm Eq}\,({\rm s})&{\bf 360000}&110000&1.6\times10^6&&\\
 \hline
\end{array}
\end{equation*}
\footnotetext[1]{The two columns show two possible inerpretations of the experiment, one with a very low $K^N$ and one with a very high $K^N$.}
\footnotetext[2]{Fields for the plateau gap density and equilibration time are left empty as no kinetic data is available.}
\caption{\label{table_values}Model parameters used in our
simulation (Fig.~\ref{fig_sim}) and their experimental values. 
The upper block shows the input parameters, the second block the
quantities that do not depend on TM concentration, i.e., the hopping
rate $r_{\rm hop}$ and the time $t_{0.01}$ ( $t_{0.001}$) when the gap
density reaches $0.01$ ($0.001$) per lattice site.  The lower three
blocks show the concentration dependent quantities for three different
TM concentrations.  These include the equilibrium gap density
$n_G^{\rm Eq}$, the average gap size $\bar g$, the plateau gap density
$n_G^0$, and the equilibration time $t^{\rm Eq}$.  Fields for the gap
density are left empty where the equilibrium state is only sparsely
covered.}
\end{table*}

\subsection*{Finite systems}

For now we have assumed infinitely long actin filaments, neglecting
all boundary effects.  However, in many cases the finite length of
actin filaments plays an important role.  For example, the length of
an actin filament in skeletal muscle is just about $1.1\,\mu{\rm m}$ (200
monomers).

\begin{figure}
\begin{center}
\includegraphics{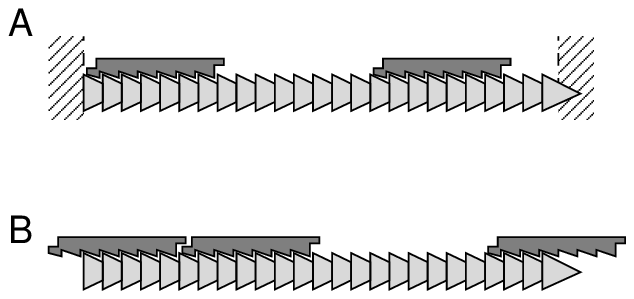}
\end{center}
\caption{Two scenarios for the boundary condition at the ends of an
  actin filament: a) a hard boundary implies that no TM molecule can
  bind beyond the end of the actin filament; b) a soft boundary allows
  TM molecules to bind to the actin end with few segments
  overhanging.}
\label{fig_boundary} 
\end{figure}

There are different possible scenarios how TM molecules should
behave at the end of an actin filament (Fig.\ \ref{fig_boundary}).  A
hard boundary would mean that a TM molecule cannot bind with any
segment overhanging the end of the actin filament.  A soft boundary,
on the other hand, would mean that binding of a TM molecule is
possible, albeit with a lower affinity, even if less than seven actin
sites are free at the end of a filament.  The binding constant of a TM
molecule partially overhanging the end of the actin filament can be
estimated in the following way.  If we neglect the entropy gain
resulting from the flexibility of the overhanging TM end the free
energy difference between the molecule bound wholly on actin and one
with $m$ overhanging segments is
\begin{equation}
\delta G_m - \delta G_0 = \frac m 7 J_0 
\end{equation}
where $J_0$ is the binding energy (including hydrophobic
contributions) of a TM molecule, which we estimate as $J_0 \approx
20\, k_B T$.  The binding constant for a TM molecule with $7-m$ bound
and $m$ overhanging segments is then
\begin{equation}
K_m = K \exp\left( - \frac{m J_0}{7 \, k_B T} \right)\;,
\end{equation}
giving the values $K_1\approx 0.057\, K$, $K_2\approx 0.0033\, K$,
etc. With the values we use ($K=10\,\mu {\rm M}^{-1}$ and $c \le
10\,\mu{\rm M}$) it turns out that only TM molecules overhanging with
one or at most two segments can bind.  We therefore conclude that
a hard boundary is a good approximation of the real situation, even if
it might not be exact.  

The process of relaxation towards equilibrium on finite filaments
differs from that on infinite ones in several aspects.  These include
the annihilation of gaps when they reach the boundary, the creation of
new gaps at the boundary and the adjustment of the whole TM block to
reduce the gaps at the boundary.  A discussion of finite-size effects
in diffusion-annihilation models can be found in \citep{krebs95}.
\begin{figure}
\begin{center}
\includegraphics{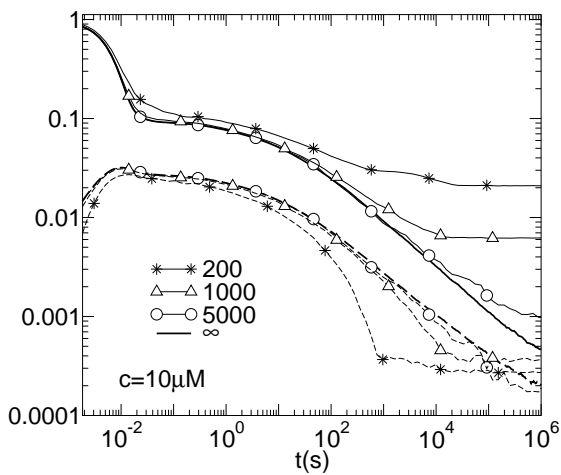}
\end{center}
\caption{Average values of the gap density ($n_G(t)$, dashed lines)
  and fraction of unoccupied binding sites ($n_0(t)\equiv 1-7 n_{\rm
    TM}(t)$, solid lines) as a function of time for filaments of
  finite length (200, 1000 and 5000 subunits), compared with results
  on ``infinite'' filaments, as obtained from a stochastic simulation.
  All parameters are the same as in Fig.\ \ref{fig_sim}, the TM
  concentration is $c=10\,\mu{\rm M}$.  The empty sites at the end of
  actin filaments are not counted as gaps, but they do contribute to
  the fraction of empty sites.  The gap density in finite systems does
  not deviate significantly from infinite systems as long as it is
  higher than one gap per filament length.}
\label{fig_finite} 
\end{figure}

Figure \ref{fig_finite} shows the gap concentration as a function of
time for different filament lengths, compared with infinitely long
filaments.  The gap concentration does not differ significantly from
infinite filaments before it reaches a value of about 1 gap per
filament length.  After that, the finite-size effects can lower (due
to faster gap annihilation at the ends) or raise (due to gap
creation at the ends) the gap concentration.

\subsection*{Time-dependent solution concentration}

In the previous sections we have assumed that the TM concentration $c$
remains constant during the experiment.  This, however, was a
simplification as the TM concentration necessarily decreases when some
of it gets bound to actin.  Situations where TM is added gradually in
course of the experiment are conceivable as well.

\begin{figure}
\begin{center}
\includegraphics{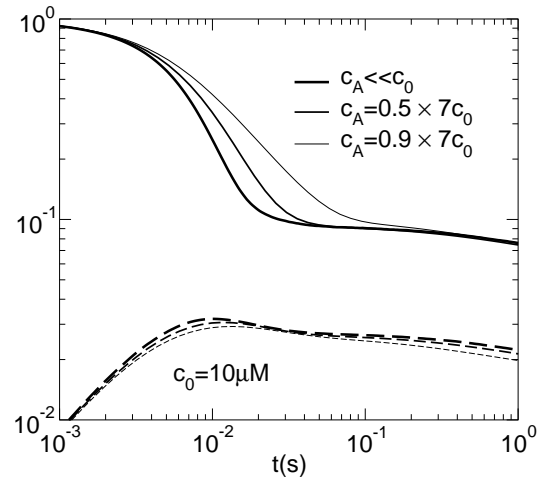}
\end{center}
\caption{Average values of the gap density ($n_G(t)$, dashed lines)
  and fraction of unoccupied binding sites ($n_0(t)$, solid lines) as
  a function of time for different actin concentrations $c_A$.  The
  simulation takes into account the drop of TM concentration as part
  of it binds to actin.  The results show that the plateau gap
  concentration is not significantly influenced by this effect.}
\label{fig_saturation} 
\end{figure}

Instead of Eq.~\ref{eq_p0} we now obtain
\begin{multline}
\label{eq_p0'}
{\bar r}_n(t)=r_n(c(t)) P_0(t) =r_n(c(t))  \times \\
\exp\left[-\int_0^t  r_n(c(t')) 
\left( \int_{t'}^t v(c(t''))dt''+2d_0 \right)dt'\right]\;.
\end{multline}
When a part of TM gets bound to actin, the solution concentration
decreases according to (the expression for $n_0(t)$ is analog to
Eq.~\ref{eq_nvt})
\begin{multline}
\label{eq_ct}
c(t)=c_0-\frac{c_A}{7}(1-n_0(t)) \\
=c_0-\frac{c_A}{7}\left(1-P_0(t)-3\int_0^t {\bar r}_n(t') dt' \right)
\end{multline}
where $c_0$ denotes the initial TM concentration and $c_A$ the
concentration of actin monomers forming the filaments. Equations
(\ref{eq_p0'}) and (\ref{eq_ct}) uniquely determine the TM and gap
concentration as a function of time and can be solved numerically.
However, in most cases the difference to the solution with a constant
concentration is not large.  This is due to the fact that the gap
concentration is determined by the number of independent nucleation
events and these mostly take place in the initial phase, when the
solution concentration has not yet dropped significantly.  An example
is shown in Fig.~\ref{fig_saturation}.  There the same curves as in
Fig.~\ref{fig_sim} are shown for the initial TM concentration of
$c_0=10\,\mu{\rm M}$ and an actin concentration which can take up
$50\%$ or $90\%$ of all TM.  Although the binding is somewhat slowed
down, the plateau gap concentration stays practically the same.  In
the second (gap annealing) stage the behavior is determined by the
detachment rates, which are independent of the concentration.  We
therefore conclude that the essential features of the system are
captured by the model with a constant TM concentration.

\section*{DISCUSSION}

Table \ref{table_values} shows the values of model parameters from the
literature and the results of our calculation for these values.  The
equilibrium gap concentration $n_G^{Eq}$ and average gap size $\bar g$
only depend on the binding constant $K^N$, the cooperativity
coefficient $\gamma$, and the TM solution concentration $c$.  For low
concentrations (with a high gap
concentration in equilibrium) the system reaches the final state very
quickly.  These situations are not the subject of our study and we
therefore leave the fields in the table empty.  We focus on cases with
a high TM coverage of actin filaments.  Interestingly, the gap
concentration after the initial binding phase $n_G^0$ only very weakly
depends on the TM concentration and other model parameters, its value
always being between 0.015 and 0.03 gaps per lattice site (between
2.8 and 5.6 gaps per micron).

The gap concentration during the annealing phase depends almost
entirely on the effective hopping rate $r_{\rm hop}$, given by
Eq. \ref{eq_hop}.  The latter is of the order of magnitude of the
smaller among the two detachment rates for a molecule at the end of a
block, $k_-^P$ and $k_-^M$.  As suggested by \citet{weigt91}, there is
strong evidence for an asymmetry in the attachment rates on the plus
($k_+^P$) and the minus end ($k_+^M$) of a block.  We therefore
estimate the hopping rate $r_{\rm hop}$ between $0.1\,{\rm s}^{-1}$
and $0.6\,{\rm s}^{-1}$.  

In most cases listed in table \ref{table_values} the equilibrium gap
concentrations were very low.  They are only relevant if the filaments
are long enough to host at least a few gaps at this low concentration.
In most experimental situations this will not be the case.  We have
shown that in a finite system the gap concentration follows the
results for an infinite system as long as the gap concentration is at
least a few per filament.  We therefore estimate the relaxation time
as the time in which the gap density drops below one per filament
length.  It scales as $t^{\rm Eq} \propto L^2$.  For a filament of
1000 lattice sites ($5.5\,\mu{\rm m}$) this gives a value between 1
and 6 hours.  The relaxation time becomes at least 100 times shorter
($40$-$220\,{\rm s}$) if one considers a state with one gap per 100
lattice sites (roughly two gaps per micron) as equilibrated. 

The amount of data gathered in different experimental studies allows
us to make quite firm predictions for the vacancy density as a
function of time.  Our modeling shows that in many relevant
experimental situations, especially when using long filaments, the
vacancy concentration is much higher than it would be in equilibrium.
Actin filaments used in skeletal muscle sarcomeres, on the other hand,
are short enough that the TM equilibrates within a minute. Therefore,
care has to be taken when the calcium regulation of myosin activity is
studied on filaments assembled in vitro, as calcium regulation is by
itself a strongly cooperative process and therefore vulnerable to gaps
in TM filaments. We also show that with realistic parameters the
effect of the initial solution concentration on the vacancy density is
rather small (not more than a factor of 2).  The only way to eliminate
gaps is to give the system enough time to equilibrate (estimated as an
hour on long filaments, though the gap concentration already reaches
quite low values after a few minutes), or to assemble the actin and
the tropomyosin filaments simultaneously.  

Unfortunately, in none of
the existing studies could the gap concentration be measured directly.
But there could be indirect ways to test the predictions of our model.
One possibility would be suddenly to decrease the TM concentration (by
dilution) or binding affinity to actin (by an increase in the salt
concentration) and study the detachment kinetics.  As it is almost
exclusively the TM molecules next to gaps that detach, the initial
detachment rate could be a direct measure for the gap concentration.

To conclude, we have shown that the dynamics of tropomyosin binding to
actin polymers shows many stages. An initial fast binding phase is
followed by a slow annealing process before it reaches the final
equilibrium.  We could give analytical approximations for the time
dependent gap density in all regimes, using elementary
nucleation-growth models for the first stage and and a mapping to a
reaction-diffusion model in the second stage.

\begin{acknowledgments}
I am grateful to Bernhard Brenner for having introduced me to the
problem of tropomyosin binding, to Tom Duke and Jaime Santos for the
careful reading of the manuscript and to Erwin Frey for many helpful
discussions.  This work was supported through a European Union
Marie Curie Fellowship (contract no.~HPMFCT-2000-00522) and partly by
the Deutsche Forschungsgemeinschaft (grant SFB-413).
\end{acknowledgments}

\end{document}